\documentclass[conference,a4paper]{APSIPA2025}
\usepackage{amsmath}
\usepackage{graphicx}
\usepackage[absolute,overlay]{textpos} 
\usepackage{subcaption}

\usepackage{multirow}
\usepackage{threeparttable}
\usepackage{booktabs}
\usepackage[backend=biber,style=ieee,]{biblatex}
\usepackage{geometry}
\usepackage{fancyhdr}

\begin{document}

    \title{DOA Estimation with Lightweight Network on LLM-Aided Simulated Acoustic Scenes}

\author{
\authorblockN{
Haowen Li\authorrefmark{1}, Zhengding Luo\authorrefmark{1}, Dongyuan Shi\authorrefmark{2}, Boxiang Wang\authorrefmark{1}, Junwei Ji\authorrefmark{1}, Ziyi Yang\authorrefmark{1}and
Woon-Seng Gan\authorrefmark{1}
}

\authorblockA{
\authorrefmark{1}
Nanyang Technological University, Singapore \\
\authorrefmark{2}
Northwestern Polytechnical University, China \\
E-mail: haowen.li@ntu.edu.sg, luoz0021@e.ntu.edu.sg, dongyuan.shi@nwpu.edu.cn,\\\{boxiang001,junwei002, ziyi016\}@e.ntu.edu.sg, ewsgan@ntu.edu.sg}}

\maketitle
\pagestyle{fancy}

\begin{abstract}
Direction-of-Arrival (DOA) estimation is critical in spatial audio and acoustic signal processing, with wide-ranging applications in real-world. Most existing DOA models are trained on synthetic data by convolving clean speech with room impulse responses (RIRs), which limits their generalizability due to constrained acoustic diversity. In this paper, we revisit DOA estimation using a recently introduced dataset constructed with the assistance of large language models (LLMs), which provides more realistic and diverse spatial audio scenes. We benchmark several representative neural-based DOA methods on this dataset and propose LightDOA, a lightweight DOA estimation model based on depthwise separable convolutions, specifically designed for mutil-channel input in varying environments. Experimental results show that LightDOA achieves satisfactory accuracy and robustness across various acoustic scenes while maintaining low computational complexity. This study not only highlights the potential of spatial audio synthesized with the assistance of LLMs in advancing robust and efficient DOA estimation research, but also highlights LightDOA as efficient solution for resource-constrained applications.

\end{abstract}

\section{Introduction}

Spatial sound perception and directional cues are crucial in real-world applications~\cite{anc+doa1,robot1,speechenhancement}. As a core technique in array signal processing, Direction-of-Arrival (DOA) estimation has recently shown great promise for enhancing learning-based ANC systems by providing spatial priors that improve noise suppression and robustness~\cite{SFANC-FxNLMSLuo, LuoBayes, DYTransfer-SFANC}.
In recent years, neural network-based and other data-driven approaches have gained popularity due to their ability to model complex spatial cues from acoustic signals. Numerous architectures have been explored for this task, including multi-layer perceptrons (MLPs)~\cite{MLP1,MLP2,MLP3}, convolutional neural networks (CNNs)~\cite{CNN1,CNN2,CNN3}, and convolutional recurrent neural networks (CRNNs)~\cite{CRNN1-BASLINE,CRNN2}, all of which have shown promising results under controlled settings.

However, these methods face a key limitation: their success heavily relies on the diversity and realism of training data. Most existing datasets are constructed by convolving clean speech with simulated room impulse responses (RIRs), resulting in acoustically constrained and semantically limited scenarios~\cite{hybrid-lhw,yyc-baseline}. Such simplified data often fails to reflect the complexities of real-world environments, where background noise, reverberation, and diverse sound events are common. As a result, models trained on these datasets tend to struggle with generalization in unseen or acoustically diverse conditions.

To better bridge the gap between synthetic and real-world acoustic conditions, we leverage a recently proposed spatial audio dataset constructed with the assistance of large language models (LLMs)-\textit{Both Ears Wide Open (BEWO)}~\cite{bewo-dataset}. Compared to traditional RIR-based corpora, this dataset introduces significantly greater diversity in both acoustic and semantic dimensions, including non-speech content, various environments, and dynamic contexts. This not only offers a more faithful proxy for real-world sound scenes, but also presents new challenges by exposing models to more complex conditions, raising the bar for network robustness and adaptability.


In this paper, we investigate single-source (SS) DOA estimation on the BEWO dataset, systematically evaluate several representative neural DOA models under this challenging setting, and propose \textit{LightDOA}, a lightweight CNN-based architecture that leverages depthwise separable convolutions to efficiently extract spatial cues from two-channel input. Inspired by the MobileNet family~\cite{mobilenet1,mobilenet2,mobilenet3,cp-mobile}, \textit{LightDOA} achieves a favorable balance between accuracy and computational cost. Experimental results show that our method achieves comparable accuracy to existing approaches while significantly reducing model complexity. Therefore, \textit{LightDOA} holds promise for deployment in downstream tasks such as ANC system~\cite{LuoRL, luoMssp}, where both source localization and real-time noise suppression are required on edge devices.

\textbf{Our contributions are threefold:}
\begin{itemize}
\item We benchmark several representative neural DOA models on a realistic and diverse two-channel dataset synthesized with LLM assistance;
\item We propose \textit{LightDOA}, a lightweight CNN-based architecture based on depthwise separable convolutions, 
\item We demonstrate that \textit{LightDOA} maintains competitive accuracy with significantly lower complexity, benefiting from the diverse spatial data generated with LLM assistance. This highlights its potential for efficient DOA estimation and deployment in downstream applications.
\end{itemize}

\section{Dataset Description} \label{sec:dataset}
\subsection{Dataset Overview}
\begin{figure*}[htbp]
    \centering
    \includegraphics[width=0.95\linewidth]{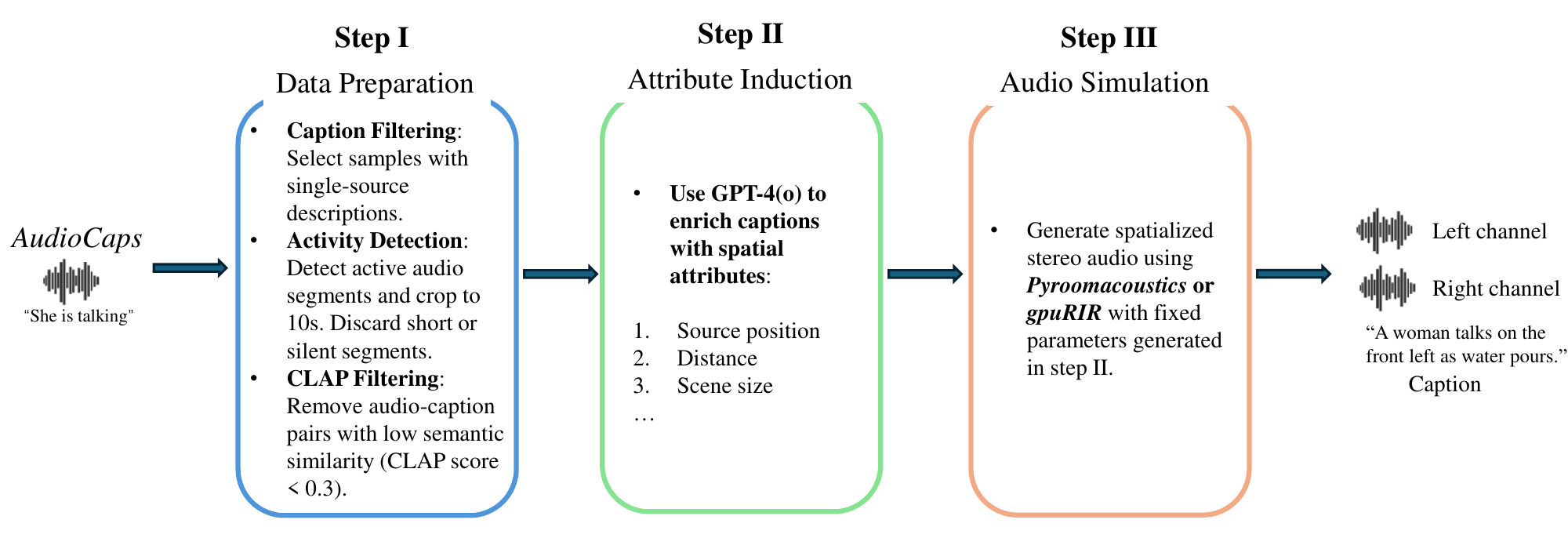}
    \caption{Pipeline for Constructing the SS Subset from \textit{AudioCaps} ~\cite{audiocaps}.}
    \label{fig:ss_pipeline}
\end{figure*}
We utilize the SS set of the BEWO dataset~\cite{bewo-dataset}~\footnotemark \footnotetext{Dataset: \url{https://huggingface.co/datasets/spw2000/BEWO-1M}}, which is a large-scale synthetic spatial audio corpus created by combining captions generated by LLMs with physically based acoustic rendering.
The process of constructing the SS-set from AudioCaps is illustrated in Fig.~\ref{fig:ss_pipeline}, including caption filtering, activity detection, CLAP~\cite{clap}-based audio-text alignment filtering, spatial attribute enrichment via GPT-4(o)~\cite{gpt4}, and stereo audio simulation using Pyroomacoustics~\cite{pyroomacoustics} or gpuRIR~\cite{gpurir}. 

\begin{table}[t]
\centering
\caption{DOA dataset composition used in this study.}
\label{tab:dataset_overview}
\begin{tabular}{|l|c|l|}
\hline
\textbf{Subset} & \textbf{Samples} & \textbf{Description} \\
\hline
Training   & 43,213 & For model training \\
Validation & 1,981  & For early stopping, tuning \\
Test       & 4,301  & For final evaluation \\
\hline
\textbf{Source}          & \multicolumn{2}{l|}{AudioCaps ~\cite{audiocaps} (caption-based retrieval)} \\
\textbf{Spatialization}  & \multicolumn{2}{l|}{Simulated with room acoustics and IPD/ILD} \\
\textbf{Content Type}    & \multicolumn{2}{l|}{Single dominant source per scene} \\
\textbf{Labels}          & \multicolumn{2}{l|}{Azimuth DOA angles} \\

\hline
\end{tabular}
\end{table}

The DOA dataset used in our study is summarized in Table~\ref{tab:dataset_overview}, including train/val/test splits, audio source, and labels. 
All audio samples in the SS-subset contain only a single dominant source per scene, ensuring spatial labels are well-defined.

\begin{figure*}[htbp]
    \centering
    \begin{subfigure}[b]{0.49\textwidth}
        \centering
        \includegraphics[width=\linewidth]{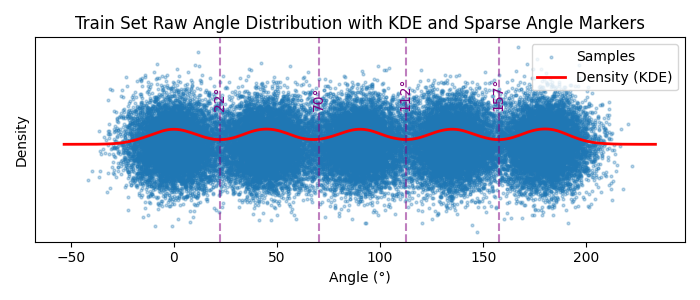}  
        \caption{KDE of training angles with sparse regions highlighted}
        \label{fig:angle_hist}
    \end{subfigure}
    \hfill
    \begin{subfigure}[b]{0.49\textwidth}
        \centering
        \includegraphics[width=\linewidth]{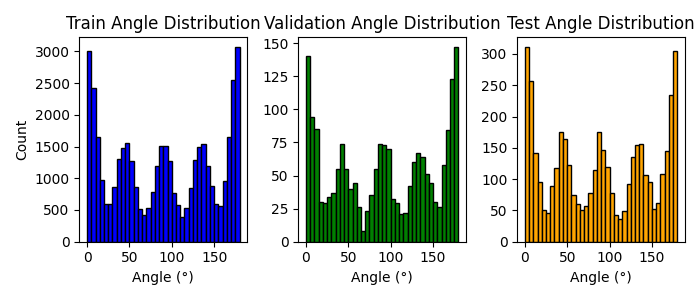}  
        \caption{Histogram of angle classes across train/val/test splits}
        \label{fig:angle_kde}
    \end{subfigure}
    \caption{Angle distribution of the BEWO dataset. 
    (a) shows the kernel density estimation (KDE) of raw DOA angles in the training set, where sparse regions (e.g., around 22°, 67°, 112°, and 157°) are visually identifiable. (b) shows the histogram of DOA angles after applying the front–back mapping across training, validation, and test sets, which are globally consistent but imbalanced within each set. }
    \label{fig:angle_imbalance}
\end{figure*}

\textbf{Limitation on Front–Back Resolution:}
Despite being dual-channel, the BEWO dataset does not incorporate head-related transfer functions (HRTFs) or any form of anatomical filtering. Consequently, directional cues are limited to interaural phase difference (IPD) and interaural level difference (ILD). As shown in Fig.\ref{fig:angle_imbalance}(a), the raw DOA angles can result in front–back ambiguity. To mitigate this, a front–back mapping is applied to fold all angles into the $[0^\circ, 180^\circ]$ range. The resulting class distributions for the training, validation, and test subsets are shown in Fig.\ref{fig:angle_imbalance}(b).

\textbf{Imbalanced Class Distribution:}
Another limitation lies in the distribution of DOA angles across the dataset. Although the train, validation, and test splits maintain similar overall statistics, the number of samples per class (i.e., per direction) is not uniformly distributed. Certain angles are overrepresented, while others have relatively few examples. This internal class imbalance can affect model training and evaluation, especially in tasks requiring fine-grained angular resolution. This imbalance is illustrated in Figure~\ref{fig:angle_imbalance}.

\subsection{Room and Microphone Configuration}
Each scene is simulated within a cuboid room $[R_0, R_1, R_2]$ based on a scene type-specific base size $r$, perturbed with uniform noise:
\begin{align}
[R_0,\; R_1,\; R_2] &= [r + \xi_{r0},\; r + \xi_{r1}, \nonumber \quad r + \xi_{r2}], \\
                    & \quad 
\xi_{ri} \sim \mathcal{U}(-0.1r,\; 0.1r)
\end{align}

Room size $r$ and reverberation time (RT60) are detailed in Table~\ref{tab:bewo-dataset}. We use $\mathcal{U}(a, b)$ to denote a uniform distribution over the range $[a, b]$.

\paragraph{Microphone Placement.} 
A stereo microphone array is positioned near the room center:
\begin{align}
[M_0,\; M_1,\; M_2] &= \left[\frac{R_0}{2} + \xi_{m0},\; \frac{R_1}{2} + \xi_{m1},\right. \nonumber \quad \left.\frac{R_2}{2} + \xi_{m2} \right],\\
                    & \quad 
\xi_{mi} \sim \mathcal{U}(-0.1r,\; 0.1r)
\end{align}
with microphones placed along axis $M_1$ using a fixed inter-microphone spacing sampled from the range $[0.16, 0.18]\mathrm{m}$, $\mathcal{U}(a, b)$ to represent a uniform distribution over $[a, b]$.

\paragraph{Source Placement.}
The source azimuth $\theta$ is sampled around class-centered angles, and converted to Cartesian coordinates based on $d$ and microphone center:
\begin{equation}
\mu_{\text{begin}} = [M_0 + d\sin{\theta},\; M_1 + d\cos{\theta},\; M_2]
\end{equation}
Class-based distributions for $\theta$ are listed in Table~\ref{tab:bewo-dataset}.
\subsection{Source Distance Modeling}
The source–array distance $d$ is sampled by applying a relative ratio $\alpha_d$ to the shortest axis-aligned distance from the microphones to the room boundaries:
\begin{equation}
d = \alpha_d \cdot \min(R_0 - M_0,\; R_1 - M_1,\; M_0,\; M_1)
\end{equation}
Here, $\alpha_d$ is selected to represent near-, moderate-, and far-field distances. See Table~\ref{tab:bewo-dataset} for detailed sampling settings.

\begin{table}[ht]
\centering
\begin{threeparttable}
\caption{Scene simulation parameters in the BEWO dataset.}
\label{tab:bewo-dataset}
\begin{tabular}{|c|c|c|}
\hline
\textbf{Attribute} & \textbf{Options List} & \textbf{Sampling Value} \\
\hline
Room size $r$ & \begin{tabular}[c]{@{}c@{}}Outdoors\\ Large\\ Moderate\\ Small\end{tabular} & 
\begin{tabular}[c]{@{}c@{}}$100\,\text{m}$\\ $\mathcal{U}(40, 90)$\\ $\mathcal{U}(20, 40)$\\ $\mathcal{U}(5, 20)$\end{tabular} \\
\hline
Direction $\theta$ & \begin{tabular}[c]{@{}c@{}}Left\\ Front-left\\ Front\\ Front-right\\ Right\end{tabular} & 
\begin{tabular}[c]{@{}c@{}}$\mathcal{N}(180^\circ,\; 11^\circ)$\\ $\mathcal{N}(135^\circ,\; 11^\circ)$\\ $\mathcal{N}(90^\circ,\; 11^\circ)$\\ $\mathcal{N}(45^\circ,\; 11^\circ)$\\ $\mathcal{N}(0^\circ,\; 11^\circ)$\end{tabular} \\
\hline
Distance ratio $\alpha_d$ & \begin{tabular}[c]{@{}c@{}}Far\\ Moderate\\ Near\end{tabular} & 
\begin{tabular}[c]{@{}c@{}}$\mathcal{U}(0.6,\; 0.9)$\\ $\mathcal{U}(0.3,\; 0.6)$\\ $\mathcal{U}(0.1,\; 0.3)$\end{tabular} \\
\hline
RT60 & — & $\mathcal{U}(0.3,\; 0.6)$ (or $0$ for outdoors) \\
\hline
Mic spacing & — & $0.16$–$0.18$\,m \\
\hline
\end{tabular}
\begin{tablenotes}
\footnotesize
\item \textit{Note:} $\mathcal{U}(a,\; b)$ indicates a uniform distribution over $[a,\; b]$, and $\mathcal{N}(\mu,\; \sigma)$ denotes a normal distribution with mean $\mu$ and standard deviation $\sigma$.
\end{tablenotes}
\end{threeparttable}
\end{table}

\section{Proposed Method}
\subsection{Feature Extraction}
We formulate the dual-channel DOA estimation task as a classification problem over discrete azimuth angles. Given a dual-channel audio waveform $(x_1(t), x_2(t))$ captured by a two-microphone array, the objective is to predict the azimuth angle $\theta \in [0^\circ, 180^\circ]$ of the dominant sound source.

We first transform the time-domain waveforms into the time-frequency domain using the Short-Time Fourier Transform (STFT):
\begin{align}
    X_1(f, t) &= \text{STFT}[x_1(t)], \\
    X_2(f, t) &= \text{STFT}[x_2(t)],
\end{align}
where $X_1(f, t)$ and $X_2(f, t)$ are the complex-valued spectrograms of the two input channels.

To capture inter-channel spatial cues, we compute the IPD, defined as the difference in phase between the two channels at each time-frequency (TF) bin:
\begin{equation}
    \text{IPD}(f, t) = \angle X_1(f, t) - \angle X_2(f, t),
\end{equation}
where $\angle X$ denotes the phase of a complex value. The resulting feature $\text{IPD}(f, t)$ is a real-valued matrix with the same shape as the input spectrogram, and serves as the input to our neural network.

\subsection{Network Architecture}

\begin{figure*}[htbp]
    \centering
    \includegraphics[width=0.99\linewidth]{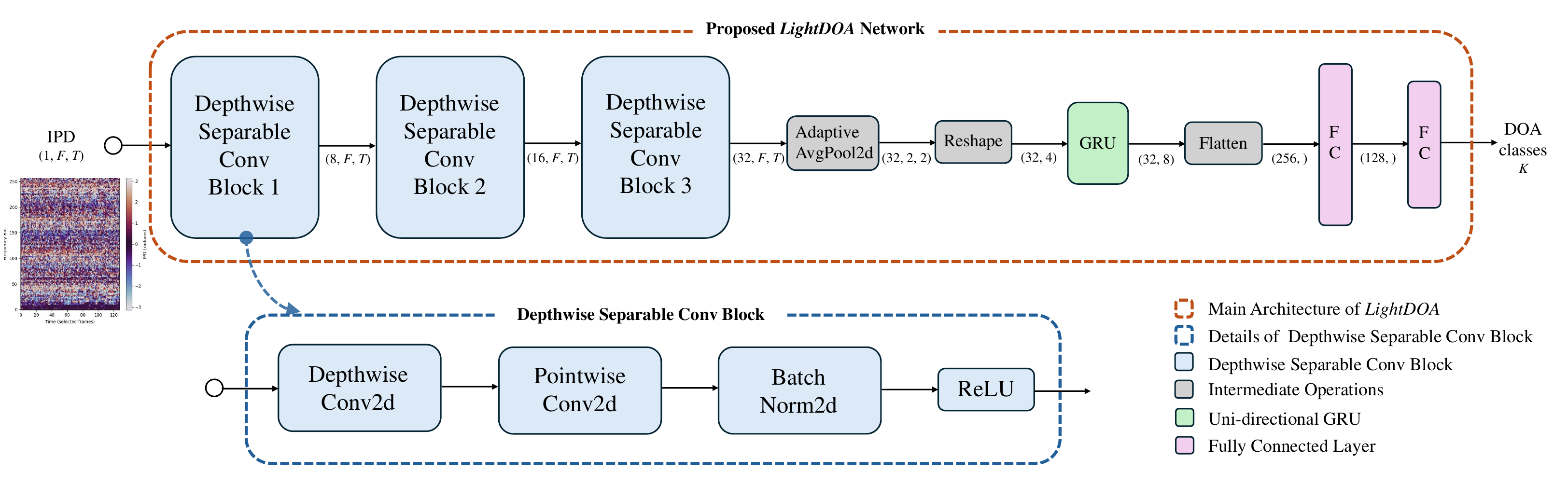}  
    \caption{ Overview of the proposed \textit{LightDOA} architecture.
The input IPD feature is processed by a stack of depthwise separable convolutional blocks, followed by temporal modeling and classification.
The region enclosed in the orange dashed line indicates the overall architecture of the proposed \textit{LightDOA} network.
The blue dashed box illustrates the internal structure of a single depthwise separable convolutional block.}
    \label{fig:model_architecture}
\end{figure*}

To efficiently model the directional information embedded in IPD, we design a lightweight neural network architecture named \textit{LightDOA}, illustrated in Figure~\ref{fig:model_architecture}. The model consists of a convolutional frontend followed by a recurrent backend, tailored for time-frequency DOA classification using two-channel input.

The input IPD feature map, with shape $(B, 1, F, T)$, is passed through a stack of depthwise separable convolutional blocks. These layers progressively extract spatial representations while maintaining low computational cost. After three convolutional stages, we apply a $2 \times 2$ adaptive average pooling to compress the feature map, yielding a compact representation of shape $(B, 32, 2, 2)$.

The pooled features are then reshaped and permuted to match the input format of a uni-directional gated recurrent unit (GRU), treating the time axis as the sequence dimension. Specifically, the GRU operates on features of shape $(B, T=32, F=4)$, producing a sequence embedding that captures temporal dependencies across frames.

The GRU output is flattened and passed through two fully connected (FC) layers to produce a 37-dimensional classification output, corresponding to azimuth classes in $[0^\circ, 180^\circ]$ with $5^\circ$ resolution. The final prediction is computed as the expected angle over the softmax probabilities, enabling smooth regression during inference.

This architecture balances representational capacity and efficiency, making it well-suited for real-time or resource-constrained spatial audio applications.

\subsection{Loss Function}
As discussed in Section~\ref{sec:dataset}, the BEWO dataset lacks HRTFs, resulting in front-back ambiguity.
To address this issue, we apply a symmetric angle mapping during preprocessing. Specifically, we first normalize all angles into the range $[0^\circ, 360^\circ)$ using:

\begin{equation}
\theta_{\text{norm}} = \theta \bmod 360^\circ
\label{eq:angle_norm}
\end{equation}

Then, we fold the full-circle range into the front-facing half-circle $[0^\circ, 180^\circ]$ as:

\begin{equation}
\theta_{\text{mapped}} =
\begin{cases}
\theta_{\text{norm}}, & \text{if } \theta_{\text{norm}} \leq 180^\circ \\
360^\circ - \theta_{\text{norm}}, & \text{if } \theta_{\text{norm}} > 180^\circ
\end{cases}
\label{eq:angle_map}
\end{equation}

This mapping ensures that front--back symmetric directions are assigned to the same target class, for example, $60^\circ$ and $300^\circ$, or $150^\circ$ and $210^\circ$.
We formulate DOA estimation as a multi-class classification task, where the half-plane is discretized into $K$ uniform angular bins. The number of classes $K$ depends on the chosen angular resolution (e.g., $K=37$ for $5^\circ$ spacing, $K=19$ for $10^\circ$, etc.). The ground-truth label $y \in \{0, 1, \ldots, K-1\}$ is assigned based on the discretized azimuth.

The model is trained using the standard cross-entropy loss~\cite{celoss}:
\begin{equation}
\mathcal{L}_{\text{CE}} = - \sum_{k=0}^{K-1} \delta_{k,y} \log p_k,
\end{equation}
where $p_k$ is the predicted probability for class $k$, and $y$ is the ground-truth class label. Note that the symmetric mapping is applied prior to training and evaluation, and is not explicitly incorporated into the loss function.

\section{EXPERIMENTS AND ANALYSES}
\subsection{Experimental Setup}

We conduct all experiments using the BEWO dataset, which simulates diverse spatial audio environments with dual-channel input. Details of the dataset generation process and spatial coverage are described in Section~\ref{sec:dataset}. Each sample contains a single dominant source with a ground-truth azimuth angle, mapped to the $[0^\circ, 180^\circ]$ range via symmetric folding.

Our experiments are conducted under four angular resolution settings, using $K \in \{9, 13, 19, 37\}$ classes corresponding to $20^\circ$, $15^\circ$, $10^\circ$, and $5^\circ$ spacing, respectively. The model is trained to minimize cross-entropy loss, and evaluated using classification accuracy in degrees.

We use the Adam optimizer~\cite{adam} with a learning rate of $5 \times 10^{-3}$, a batch size of 256, and train for up to 150 epochs. Early stopping is applied based on the validation accuracy: training is terminated if no improvement is observed over 10 consecutive epochs. No data augmentation is used.

Model selection is based on the checkpoint that achieves the highest validation accuracy. All reported results are evaluated on the test set using this best-performing checkpoint. Each experiment is repeated with three random seeds to ensure result stability.
\subsection{Baselines}

To evaluate the effectiveness of our proposed \textit{LightDOA} model, we compare it against three representative baseline methods, each reflecting a different design philosophy in neural-based DOA estimation:
\begin{table*}[htbp]
\centering
\caption{Classification accuracy (\%) and model parameters under different angular resolutions. Parameters vary across resolutions due to different numbers of output classes.}
\label{tab:doa_accuracy_detailed}
\resizebox{\textwidth}{!}{
\begin{tabular}{l|l|cc|cc|cc|cc}
\toprule
\textbf{Model} & \textbf{Feature} & 
\multicolumn{2}{c|}{\textbf{5°}} & 
\multicolumn{2}{c|}{\textbf{10°}} & 
\multicolumn{2}{c|}{\textbf{15°}} & 
\multicolumn{2}{c}{\textbf{20°}} \\
& & Acc (\%) & Param & Acc (\%) & Param & Acc (\%) & Param & Acc (\%) & Param \\
\midrule
CRNN~\cite{CRNN1-BASLINE}          & STFT (Re+Im)           & 57.29 & 311k & 71.08 & 309k & 76.61 & 308k & 84.26 & 308k \\
MTL-DOA~\cite{yyc-baseline}               & IPD+logMag     & 56.68 & 285m & 70.24 & 285m & 78.15 & 285m & 85.00 & 285m \\
CP-Mobile~\cite{cp-mobile}       & IPD            & 57.48 & 64.0k & 71.03 & 62.1k & 78.14 & 61.5k & 84.40 & 61.1k \\
CP-Mobile~\cite{cp-mobile}     & STFT (Re+Im)           & 54.48 & 64.2k & 69.05 & 62.3k & 76.68 & 61.7k & 85.00 & 61.4k \\
\textbf{Proposed \textit{LightDOA} } & \textbf{IPD} & \textbf{57.96} & \textbf{39.0k} & \textbf{71.66} & \textbf{36.7k} & \textbf{77.98} & \textbf{35.9k} & \textbf{85.45} & \textbf{35.5k} \\
\bottomrule
\end{tabular}}
\vspace{0.3em}
\caption*{\textit{Note:} The baseline models were re-implemented and slightly adjusted to fit the current DOA estimation task. In particular, their output layers were adapted to match the number of DOA classes under each angular resolution setting. Other parts of the architecture remained unchanged.}
\end{table*}
\begin{itemize}
    \item \textbf{CRNN}~\cite{CRNN1-BASLINE}: A popular end-to-end neural network for DOA estimation, originally designed to jointly estimate azimuth and elevation angles using acoustic intensity vectors as input~\cite{CRNN1-BASLINE}. Since our dataset provides only dual-channel recordings, we adapt the model to predict azimuth only and replace the original input with the input consists of the real and imaginary parts of the complex STFT, concatenated along the channel dimension, which called STFT (Re+Im) in table~\ref{tab:doa_accuracy_detailed}. This baseline serves as a strong full-capacity reference using spectral features.

    \item \textbf{MTL-DOA}~\cite{yyc-baseline}: A multi-task learning framework that uses separate branches to process IPD and log-magnitude spectrograms (logMag), followed by a joint fusion module to estimate azimuth and elevation~\cite{yyc-baseline}. In our setting, we evaluate only the azimuth estimation performance, adapting the model to our dual-channel setup. This baseline reflects the benefit of multi-view spatial representation learning.

    \item \textbf{CP-Mobile}~\cite{cp-mobile}: To assess performance–efficiency trade-offs, we adapt the compact CP-MobileNet~\cite{cp-mobile}, widely used in acoustic scene classification, for DOA classification. The model is tested with both IPD and STFT (Re+Im) inputs, and its output layer is modified to predict azimuth angle classes. This baseline provides insight into the potential of lightweight architectures for DOA tasks.
\end{itemize}

These baselines cover a diverse spectrum of design choices, from full-capacity CRNNs to efficient mobile models and multi-branch feature encoders. All models are trained and evaluated under the same experimental conditions as \textit{LightDOA} to ensure fair comparison.

\subsection{Results and Discussion}


Table~\ref{tab:doa_accuracy_detailed} presents the classification accuracy and corresponding parameter counts of various DOA estimation models under different angular resolutions (5°, 10°, 15°, and 20°). The performance is evaluated on the test set using the best validation checkpoint, as described in Section~4.1.
Since the number of output classes varies with the angular resolution (e.g., 37 classes for 5°, 9 classes for 20°), the output layers of the models are adjusted accordingly.
However, these changes have minimal impact on the overall parameter count, and the complexity remains relatively stable across different configurations, ensuring fair comparison among models.

Among all methods, our proposed \textit{LightDOA} model consistently achieves the highest accuracy across all resolution settings, with performance gains especially notable at coarser resolutions (e.g., 85.45\% at 20°). Despite its lightweight design with only (or less than) 39k parameters, it outperforms larger models such as CRNN (around 311k)~\cite{CRNN1-BASLINE} and the MTL-DOA model (around 285m)~\cite{yyc-baseline}, indicating superior efficiency–accuracy trade-off.

In comparison, CRNN-based model\cite{CRNN1-BASLINE} perform competitively but exhibit lower accuracy at finer resolutions, and their relatively large parameter size may limit deployment in real-time or embedded applications. CP-Mobile variants~\cite{cp-mobile}, adapted from acoustic scene classification tasks, also demonstrate reasonable accuracy with moderate complexity, but underperform compared to our proposed model, particularly when using STFT(Re+Im) input.

Notably, the performance drop observed in baseline methods relative to their original results is likely due to the greater diversity and inherent class imbalance in BEWO dataset as shown in Section~\ref{sec:dataset}, which poses a more challenging evaluation setting.

These results demonstrate that our \textit{LightDOA} architecture not only achieves state-of-the-art accuracy, but also sets a new standard for low-complexity DOA estimation. The results validate the effectiveness of IPD-based input and depthwise separable convolution for compact spatial audio modeling.

\section{Conclusions}
In this work, we revisited single-source DOA estimation using a recently introduced dual-channel dataset BEWO, where LLMs were used to assist spatial audio generation. To address the limitations of existing models in terms of generalizability and complexity, we proposed \textit{LightDOA}, a lightweight neural network architecture based on depthwise separable convolutions and IPD features.
Extensive experiments demonstrated that \textit{LightDOA} achieves competitive or superior accuracy compared to existing CRNN-based model, multi-branch architectures and CP-Mobile variants, while maintaining significantly lower model complexity. The lightweight nature of our model makes it a promising candidate for real-time deployment on edge devices.
Future work may explore extensions to elevation estimation, multi-source localization, or adaptation to real-world recordings beyond synthetic datasets.











\begingroup
\small  
\printbibliography
\endgroup
\end{document}